\documentclass[twocolumn,showpacs,preprintnumbers,draftclsnofoot]{revtex4}
\usepackage{amsmath}
\usepackage{amssymb}
\usepackage{graphicx}
\usepackage{dcolumn}
\usepackage{bm}
\usepackage{amssymb}
\usepackage{graphicx}
\usepackage{dcolumn}
\usepackage{bm}

\begin{document}

\title{ Valley selecting current partition at zero-line mode of quantum anomalous Hall topologies  }
\author{Ma Luo\footnote{Corresponding author:luom28@mail.sysu.edu.cn} }
\affiliation{The State Key Laboratory of Optoelectronic Materials and Technologies \\
School of Physics\\
Sun Yat-Sen University, Guangzhou, 510275, P.R. China}

\begin{abstract}

Topologically protected zero-line modes appear at the interface between two regions of the monolayer graphene in quantum anomalous Hall phase with different Chern number. In the presence of staggered sublattice potential, the band gaps of the two valleys become different, and the phase diagram defined by the Chern number has an additional regime of topologically trivial phase. The interface between the topologically trivial and non-trivial regions hosts zero-line mode in only one valley. By tuning the exchange field, three types of interface that host zero-line modes in selected valley(s) are formed. The nano-devices consisted of Y-shape junctions of the three types of interface exhibit the functions of valley splitting, merging or filtering for the incident currents.

\end{abstract}

\pacs{0000000000000000} \maketitle

\section{Introduction}

Investigation of the quantum anomalous Hall(QAH) effect in two dimensional(2D) materials has been attracting vast interest\cite{YafeiRen16r,HMWeng15r,CXLiu16r,RuiYu10,ZhenhuaQiao10,WangKongTse11,HongbinZhang12,ZFWang13,ShuChunWu14,ZhenhuaQiao14,GangXu15,JLLado15,ShifeiQi16,JianZhou16,LiangDong16}, because of the novel topological properties \cite{YafeiRen16r,HMWeng15r,CXLiu16r}. Monolayer graphenes in the presence of the Rashba spin-orbital coupling (SOC) and the exchange field exhibit QAH phase\cite{ZhenhuaQiao10,ZhenhuaQiao14}. The quantized charge Hall conductance is determined by the Chern number, which label the global property of the band structure. Because of the topological non-trivial property of the band structure, at the edge of the graphene nano-ribbon, robust gapless edge states appear. Another type of edge states appear at the interface between two topologically non-trivial insulators with opposite topological invariants, which is call the zero-line mode(ZLM)\cite{Ivar08,Matthew11,Zarenia11,Klinovaja12,Abolhassan13,FanZhang13,XintaoBi15,ChangheeLee16}. The ZLMs is localized near to the interface and propagating along the interface. Current partition at the junction of four ZLMs splits the incident current into two currents\cite{ZhenhuaQiao11,ZhenhuaQiao14P,Anglin17,KeWang17,YafeiRen17}. The prediction of the optical excitation of localized spin and valley currents promote the ZLMs as promising candidate for integrated optical spintronic and valleytronic devices\cite{luoma17}. Progress toward experimental implementation of the ZLMs has been made\cite{Augustinus12,JingLi16}. This article focus on the ZLM at the interface between two regions of the monolayer graphene in QAH phase with different Chern number \cite{YafeiRen17}. The dispersions of the ZLMs at K and K$^{\prime}$ valleys are the same, due to the breaking of the time reversal symmetric by the exchange field.

The presence of staggered sublattice potential in graphene open a band gap. This potential could be realized by h-BN \cite{Giovannetti07,CRDean10,Sachs11,Eryin16} or SiC \cite{SYZhou07,Agrawal13} substrates. We combine the staggered sublattice potential with the monolayer graphene in QAH phase. The phase diagram includes the QAH phase with Chern number being $\mathcal{C}=\pm4$ and the trivial phase with $\mathcal{C}=0$. The phase boundaries are determined by the ratio between the strength of the staggered sublattice potential and exchange field. The ZLMs at the interface between topological trivial and non-trivial phase is investigated. Because of the diversification of the phase diagram, Y-shape junction between three ZLMs is allowed. The Y-shape junction exhibit valley selecting current partition, which can be used to construct valley splitting, merging or filtering devices.


The article is organized as following: In section II, the tight binding model for the graphene with Rashba SOC, exchange field and staggered sublattice potential is given. The phase diagram and band structure of the bulk graphene is discussed in this section. The band structure of the ZLMs in the middle of the zigzag nano-ribbon is calculated. In section III, the Y-shape junctions are proposed. The two devices with valley splitting(merging) and filtering effects are simulated by applying the non-equilibrium Green's function method. In section IV, the conclusion is given.

\section{quantum anomalous Hall topologies with staggered sublattice potential}

The tight binding Hamiltonian of the graphene with exchange field, staggered sublattice potential and Rashba spin-orbital coupling(SOC) is given as
\begin{eqnarray}
H&=&-t\sum_{\langle i,j\rangle,\alpha}{c_{i\alpha}^{+}c_{j\alpha}}\nonumber \\&+&ig_{R}\sum_{\langle i,j\rangle,\alpha,\beta}{c_{i\alpha}^{+}[(\mathbf{S}\times\mathbf{d}_{ij})\cdot\hat{z}]_{\alpha\beta}c_{j\beta}}\nonumber \\&+&\lambda\sum_{i,\alpha}{c_{i\alpha}^{+}[\sigma_{z}]_{\alpha\alpha}c_{i\alpha}}+\Delta\sum_{i,\alpha}{\delta_{i}c_{i\alpha}^{+}c_{i\alpha}}
\label{Hamiltonian}
\end{eqnarray}
where i and j label the lattice sites, $\alpha=\pm1$ and $\beta=\pm1$ label the spin, the summations with index $\langle i,j\rangle$ run through the nearest neighboring sites, $t=2.8$ eV is the nearest neighbor hopping energy, $g_{R}$ is the Rashba SOC strength, $\lambda$ is the exchange field strength, $\Delta$ is the staggered sublattice potential, $\delta_{i}=+1(-1)$ for $i$ in A(B) sublattice site, $\mathbf{S}=[\sigma_{x},\sigma_{y},\sigma_{z}]$ is vector of Pauli matrixes, $\mathbf{d}_{ij}$ is the unit vector from lattice $i$ to $j$.

\begin{figure}[tbp]
\scalebox{0.58}{\includegraphics{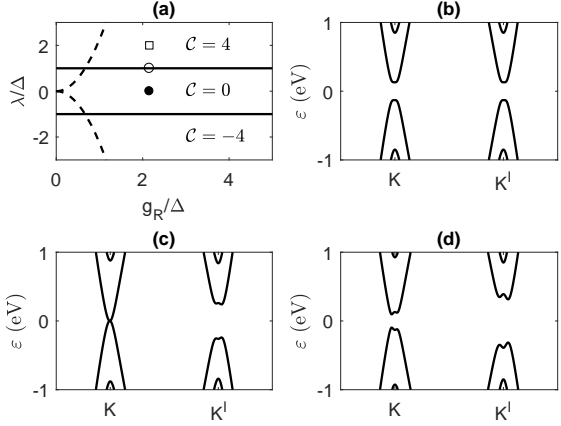}}
\caption{ (a) The phase diagram of the graphene in the $\lambda/\Delta$-$g_{R}/\Delta$ space. The solid line separate different phase, and the dash lines separate regimes with different type of band gap at K and K$^{\prime}$ points. (b) (c) and (d) are the bulk band structure of the graphene in the phases labeled by the filled circle, empty circle and empty square in (a), respectively.  }
\label{fig_bulk}
\end{figure}

\begin{figure}[tbp]
\scalebox{0.56}{\includegraphics{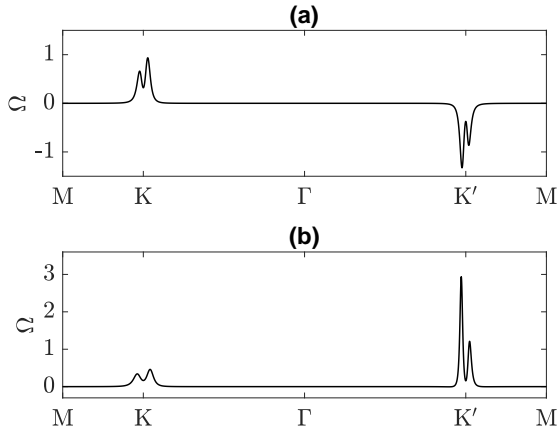}}
\caption{ The Berry curvature $\Omega$ in the unit of $nm^{2}$ along the $K-\Gamma-K^{\prime}$ line in the Brillouin zone, for the graphene with $\lambda/\Delta=0.2$ in (a) and $\lambda/\Delta=1.8$ in (b).  }
\label{fig_Berry}
\end{figure}

Applying the Bloch periodic boundary condition, the bulk band structure is obtained. The topological phase is determined by the Chern number(denoted as $\mathcal{C}$), which is the integral of the Berry curvature through the whole Brillouin zone. The phase diagram is plotted in Fig. \ref{fig_bulk}(a). The phase transition boundaries are featured by gap closing, which occur at K or K$^{\prime}$ point. In the simultaneous presence of the staggered sublattice potential and exchange field, the band structures in K and K$^{\prime}$ valley are different. Specifically, the eigenvalues at K(K$^{\prime}$) point are $\pm[\Delta+(-)\lambda]$ and $\pm\sqrt{g_{R}^{2}+(\Delta-(+)\lambda)^{2}}$. In the regime with $\lambda>(<)+(-)9g_{R}^{2}/(4\Delta)$(to the left of the dash line in Fig. \ref{fig_bulk}(a)), the band gap at K(K$^{\prime}$) point is $2\sqrt{g_{R}^{2}+(\Delta-(+)\lambda)^{2}}$; in the other regime, the band gap at K(K$^{\prime}$) point is $2[\Delta+(-)\lambda]$. Thus, the band gap at K(K$^{\prime}$) point is closed when $\lambda=-(+)\Delta$ for arbitrary $g_{R}$, or when $g_{R}=0$ and $\lambda=+(-)\Delta$. The global band gap is close at two solid lines in Fig. \ref{fig_bulk}(a), which separate the topological trivial and non-trivial phases.

The band structures of three typical systems with parameter being labeled in Fig. \ref{fig_bulk}(a) are plotted in \ref{fig_bulk}(b), (c) and (d), respectively. In the trivial phase with $\lambda=0$, the band structures of K and K$^{\prime}$ valleys are the same, as shown in Fig. \ref{fig_bulk}(b). The band gap is $2\Delta$. Increasing $\lambda$ drives the band gap in K(K$^{\prime}$) point smaller(larger). The Berry curvature of a system in this phase regime is plotted in Fig. \ref{fig_Berry}(a). The Berry curvature in K and K$^{\prime}$ valley have opposite sign. The integral of the Berry curvature through the whole Brillouin zone gives $\mathcal{C}=0$. With $\lambda=\Delta$, the gap at K point is closed, as shown in Fig. \ref{fig_bulk}(c). By contrast, with $\lambda=-\Delta$, the gap at K$^{\prime}$ point is closed. This critical phase is valley-polarized semi-metal. Further increasing $\lambda$ reopen the gap at K point. The Berry curvature of a system in this phase regime is plotted in Fig. \ref{fig_Berry}(b). In this case, the Berry curvature in K and K$^{\prime}$ valley have the same sign. The integral of the Berry curvature through half of the Brillouin zone that cover one valley gives fractional number. The integral of the Berry curvature through the whole Brillouin zone gives $\mathcal{C}=\pm4$. With $\lambda=2\Delta$, the band gap at K point reaches $2\Delta$ again, as shown in Fig. \ref{fig_bulk}(d). Note that the global band gap is slightly smaller than $2\Delta$, because the band minimal locates beyond the K point. Flipping the sign of the exchange field exchange the band structures at K and K$^{\prime}$ points. As a result, the electron transportation and optical excitation in these graphenes exhibit valley selecting property that is controlled by the exchange field.

\begin{figure}[tbp]
\scalebox{0.5}{\includegraphics{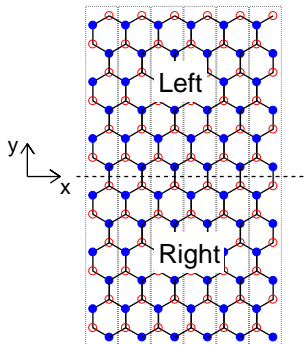}}
\caption{ Lattice structure of the zigzag nanoribbons that are periodic along the x axis. The A and B sublattice with staggered sublattice potential $\Delta$ and $-\Delta$ are plotted as empty(red) and filled(blue) dots, respectively. The domain wall is marked by the dashes line. The unit cells are marked by the dotted lines.  }
\label{fig_config}
\end{figure}

\begin{figure}[tbp]
\scalebox{0.34}{\includegraphics{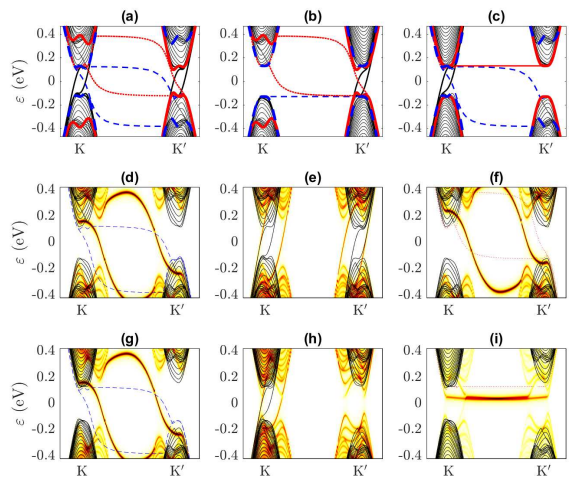}}
\caption{ Band structure of zigzag-edge nano-ribbon with $g_{R}=0.2t$ and $\Delta=130meV$. The exchange field at the left and right half of the nano-ribbon is $\lambda=+2\Delta$ and $\lambda=-2\Delta$ in (a), $\lambda=+2\Delta$ and $\lambda=0$ in (b), $\lambda=0$ and $\lambda=-2\Delta$ in (c). The total width of the nano-ribbon is 51.12 nm. The bands of the ZLMs are plotted as black(solid) lines, the bands of the chiral edge states at the left open edge are plotted as red(dotted) lines and the bands of the chiral edge states at the right open edge are plotted as blue(dashed) lines. The thick red(solid) and blue(dashed) lines are the bulk band edge of the left and right regions, respectively. The thin black(solid) lines are bulk states. In the presence of the Hubbard interaction, the spectral function of the system with parameters in (a) are plotted in (d-f). Specifically, the summations of the spectral function around the left open edge, the domain wall and the right open edge are plotted in (d), (e) and (f), respectively. The non-interacting band structures of the corresponding chiral edge states or ZLMs are plotted as thin lines for comparison. Spectral functions of the same scheme are plotted in (g-i) for the system with parameters in (c). }
\label{fig_ribbon}
\end{figure}

The ZLMs are formed at the interface between two regions with different Chern numbers. In order to exhibit the ZLMs numerically, we calculate the band structures of zigzag nano-ribbons. The lattice structure of the nano-ribbon is plotted in Fig. \ref{fig_config}. The domain wall is located in the middle of the nano-ribbon along the longitudinal direction (x axis). When the exchange fields to the left and right of the domain wall are $\lambda=+2\Delta$ and $\lambda=-2\Delta$, respectively, the band structure is shown in Fig. \ref{fig_ribbon}(a). The ZLMs at the two valleys have the same dispersion, so that the localized conductivity of the ZLMs is finite(zero) under forward(backward) bias, assuming that the Fermi level is in the bulk gap and crosses the dispersion of the ZLMs. In addition to the ZLMs, the chiral edge states localized at the left and right open edges are also gapless. The dispersion of the chiral edge state is opposite to the dispersion of the ZLMs. If the y coordinate of the domain wall is changed by $\frac{3N}{4}a_{c}$ with $a_{c}$ being the bond length of graphene and $N\in\{\pm1,\pm2,\pm3\}$, the domain wall cuts through different types of bond. The dispersions of the ZLMs would be slightly changed, but the topological features remain the same. If the domain to the right of the domain wall has zero exchange field, the ZLMs in the K valley disappear, as shown in Fig. \ref{fig_ribbon}(b). Meanwhile the edge states localized at the right open edges become two-fold degenerated flat band. The energy of the flat band is equal to the local potential of the edge atoms. By contrast, if the domain to the left of the domain wall have zero exchange field, the ZLMs in the K$^{\prime}$ valley disappear, as shown in Fig. \ref{fig_ribbon}(c). Thus, the numerical results of the band structures confirm that the number of gapless edge states at the interface between two regions is determined by the difference between the Chern numbers of the two adjacent regions.

In the presence of electron-electron interaction, the band gap would be modified, but the topological number remains the same in this case. The interaction is described by the additional term of Hubbard model, $U\sum_{i}\hat{n}_{i,\alpha}\hat{n}_{i,\bar{\alpha}}$, in the Hamiltonian of the tight binding model. In our calculation, the realistic parameter with $U=1.6t$ is used. The model can be solved by the cluster perturbation theory(CPT) method\cite{Senechal00,Senechal02,Grandi15,Grandi151,MaLuo18}, which gives the spectral function. The spectral function is the LDOS at each lattice site versus the energy and wavenumber. Summation of the spectral function at the lattice sites around the domain wall or the open edges shows the quasi-particle dispersion of the corresponding edge states. For the nano-ribbons with parameter in Fig. \ref{fig_ribbon}(a), the summations of the spectral function around the left open edge, the domain wall and the right open edge are plotted in  Fig. \ref{fig_ribbon}(d), (e) and (f), respectively. The band structures of the chiral edge states and ZLMs without interaction is plotted in the corresponding figures for comparison. In Fig. \ref{fig_ribbon}(d) and (f), within the energy range of the bulk gap, one can find that the dispersions of the chiral edge states are not significantly modified by the interaction. In Fig. \ref{fig_ribbon}(e), one can find that the momentum difference between the two ZLMs in the same valley is enlarged due to the interaction. For the nano-ribbon with parameter in Fig. \ref{fig_ribbon}(c), the summations of the spectral function around the left open edge, the domain wall and the right open edge are plotted in  Fig. \ref{fig_ribbon}(g), (h) and (i), respectively. In Fig. \ref{fig_ribbon}(i), one can find that the flat bands at the open edge of the region with $\mathcal{C}=0$ is moved toward $\varepsilon=0$ due to the interaction. The flat band become slightly dispersive with finite conductivity under forward and backward bias. Because the Hubbard interaction does not qualitatively change the conductive properties around $\varepsilon=0$, the non-interacting model is used in the following sections.


\section{valley splitting current partition}

Because changing the exchange field drives the graphene into three topological phase with different Chern number, the interfaces between any two phases support three types of ZLM. The Y-shape junction of three ZLMs has the function of current partition. We denote the three type of ZLMs in Fig. \ref{fig_ribbon}(a), (b) and (c) as ZLM-a, ZLM-b and ZLM-c.

\begin{figure}[tbp]
\scalebox{0.16}{\includegraphics{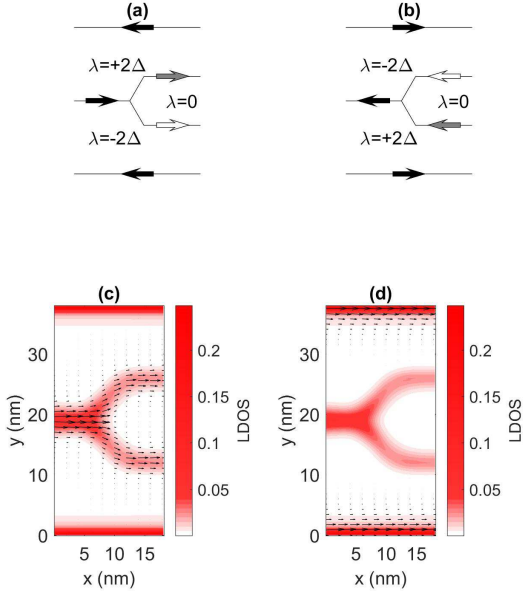}}
\caption{ (a) and (b) are spatial blueprint of the Y-shape junction in a nano-ribbon. The exchange fields of each region in (a) and (b) are labeled in the figures. The black arrow indicate the localized edge states(ZLMs or chiral edge states) that support current in both valley, the white(grey) arrow indicate the localized edge states that support valley polarized current in K(K$^{\prime}$) valley. (c) and (d) are the numerical result of the transportation calculation of the systems in (a) and (b) under forward bias, respectively. The color scale indicates the LDOS, and the vector fields indicates the local current distribution.  }
\label{fig_split}
\end{figure}


The nano-structure with valley splitting effect is presented in Fig. \ref{fig_split}(a).  Because the ZLM-b and ZLM-c only support current in K and K$^{\prime}$ valley, respectively, under forward bias, the incident charge current from the ZLM-a are split into two currents. Because all ZLMs have positive dispersion, the backward bias does not excite current in the ZLMs, but excite currents along the open edges. Flipping the sign of the exchange field at all regions gives the nano-structure with valley merging effect, as shown in Fig. \ref{fig_split}(b). In this case, the forward bias excite currents along the open edges; the backward bias excite valley polarized current at ZLM-b and ZLM-c, which merge at the Y-shape junction.

The proposal are confirmed by numerical calculation of quantum transportation based on the non-equilibrium Green's function(NEGF) theory\cite{LopezSancho84,Nardelli98,FufangXu07,Diniz12,Lewenkopf13}. The recursive algorithm is applied to construct the Green's function. The local density of state(LDOS) is given by the imaginary part of the retarded Green's function as $-\frac{1}{2\pi}Im[G^{r}(\mathbf{r})]$. The local current is given as
\begin{equation}
\mathbf{j}(\mathbf{r}_{i}\rightarrow\mathbf{r}_{j})=-\frac{2e\hat{\mathbf{d}}_{ij}}{h}\int dE[t_{ij}G^{<}(\mathbf{r}_{j},\mathbf{r}_{i})-t_{ji}G^{<}(\mathbf{r}_{i},\mathbf{r}_{j})]
\end{equation}
where $\mathbf{r}_{i}$ is the position of the i-th lattice site, $\hat{\mathbf{d}}_{ij}$ is the unit vector from the i-th to j-th lattice site, $e$ is the electron charge, $h$ is the Planck constant, $t_{ij}$ is the hopping parameter between the i-th and j-th lattice site, and $G^{<}(\mathbf{r}_{i},\mathbf{r}_{j})$ is the lesser Green's function. For the system in Fig. \ref{fig_split}, the front and back leads are the nano-ribbons with one and two domain walls, respectively; the Y-shape junction is the scattering region. The numerical results are plotted in Fig. \ref{fig_split}(c) and (d) for the Y-shape junction in Fig. \ref{fig_split}(a) and (b), respectively. Both figures present the results with forward bias. Under backward bias, the current distributions of the system in Fig. \ref{fig_split}(a) and (b) are given by the result in Fig. \ref{fig_split}(d) and (c) with reversing current direction, respectively. The results show that the system in Fig. \ref{fig_split}(a) has valley splitting effect under forward bias. In contrary, the system in Fig. \ref{fig_split}(b) has valley merging effect under backward bias.

\begin{figure}[tbp]
\scalebox{0.126}{\includegraphics{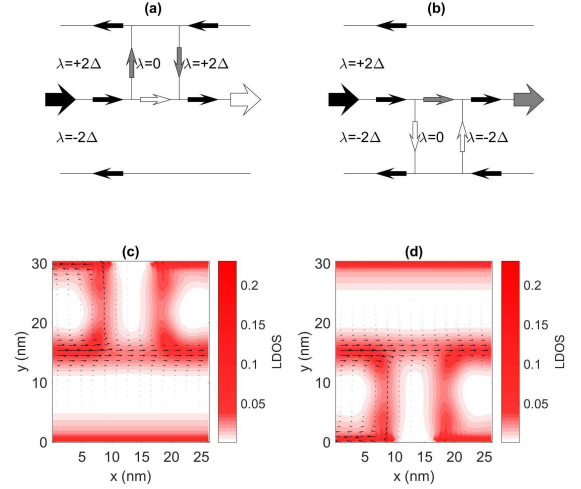}}
\caption{ The same type of plotting as in Fig. \ref{fig_split} for the valley filtering devices. In (a) and (b), the wider black arrow indicate the incident charge current; the wider white(grey) arrow indicate the exiting valley polarized current at K(K$^{\prime}$) valley at the ZLMs that support current in both valley.  }
\label{fig_filter}
\end{figure}

Although the ZLM-a support current in both valley, excitation of valley polarized current in ZLM-a could be useful for valleytronic devices. The valley filtering devices consisting of two Y-shape junction is presented in Fig. \ref{fig_filter}(a) and (b). The background of the nano-structure is the nano-ribbon with ZLM-a. In the scattering region, the exchange field in a square panel is switched to $\lambda=0$. In the system being plotted in Fig. \ref{fig_filter}(a) or (b), the square panel covers the region with $\lambda=+2\Delta$ or $\lambda=-2\Delta$, respectively. Because the LDOS around $\varepsilon=0$ at the open edge of the square panel is near to zero, this section of the open edge does not support current. For the system in Fig. \ref{fig_filter}(a), under forward bias, the incident port excite current along the ZLM-a in both of the K and K$^{\prime}$ valleys. At the first Y-shape junction, the K valley current is partitioned into the ZLM-c(parallel to the nano-ribbon) in the scattering region, while the K$^{\prime}$ valley current is partitioned into the ZLM-b(perpendicular to the nano-ribbon). At the second Y-shape junction, the K valley current transmits into the ZLM-a of the exiting port. The K$^{\prime}$ valley current is redirected into the open edge and flow back to the incident port. The devices effectively transmit the K valley current and reflect the K$^{\prime}$ valley current, which form the K valley filter. Similarly, the system in Fig. \ref{fig_filter}(b) form the K$^{\prime}$ valley filter. The proposal is confirmed by the quantum transportation calculation as well, which is plotted in Fig. \ref{fig_filter}(c) and (d).

\section{conclusion}

In summary, the ZLMs at the domain walls of monolayer graphene are studied. If the domain wall separates two regions in QAH phase with opposite Chern number, the ZLMs support one-way current in both K and K$^{\prime}$ valleys; if the domain wall separates the region in QAH phase from the region in topologically trivial phase, the ZLMs support one-way current in either K or K$^{\prime}$ valley. The Y-shape current partition at the junction of three different type of ZLMs are proposed. The devices with valley splitting, merging and filtering effects are designed. The valley filtering device makes integrated valleytronic feasible.


\begin{acknowledgments}
The project is supported by the National Natural Science Foundation of China (Grant:
11704419), the National Basic Research
Program of China (Grant: 2013CB933601), and the National Key Research and Development Project of China
(Grant: 2016YFA0202001).
\end{acknowledgments}

\clearpage

\end{document}